\documentclass{JINST}
\usepackage{graphicx}
\usepackage{mathptmx}
\usepackage{amssymb}
\usepackage{fontenc}
\usepackage{times}
\title{New gas mixtures suitable for rare event detection
using a Micromegas-TPC detector}

\author{L. Ounalli$^a$\thanks{Corresponding author.}~, J-L. Vuilleumier$^{a,b}$, D.
Schenker$^a$ and J-M. Vuilleumier$^{a,b}$\\ \llap{$^a$}Laboratory
for Particle Physics,\\ A.L. Breguet 1, CH-2000 Neuch\^atel\\
\llap{$^b$}Laboratory for High Energy Physics,\\ Sidlerstrasse 5, CH-3012 Berne\\
E-mail:\email{leila.ounalli@unine.ch}}

\abstract{The aim of the presented work was to develop further
techniques based on a Micromegas-TPC, in order to reach a high gas
gain with good energy resolution, and to search for gas mixtures
suitable for rare event detection. This paper focuses on Xenon,
which is convenient for the search of neutrinoless double beta
decay in $^{136}$Xe, and CF$_4$, suitable for dark matter searches
and the study of solar and reactor neutrinos. Various
configurations of the Micromegas plane were investigated and are
described. Gains of $10^{5}$ and energy resolutions of 35-65$\%$
at 6 keV have been achieved.}


\begin{document}

\section{Introduction}
Low energy neutrino experiments, whether at reactors \cite{MUNU2}
or with solar neutrinos \cite{Giunti}, and in particular searches
for double beta decay \cite{EXO} require high background
suppression. Good event signature can contribute significantly.
TPCs, thanks to their spatial resolution, allow a good event
selection using the event topology. Good energy resolution, of
order 1$\%$ at 2 MeV or better, is mandatory to separate
neutrinoless double beta decay "$0\nu\beta\beta$" from the allowed
two neutrino decay "$2\nu\beta\beta$". High gas amplification is
necessary for optimal performance. The Micromegas technique is
highly promising. In addition, Micromegas are found to be
reliable, and suited for large size devices. We have conducted
tests of Micromegas in various gas mixtures of interest for low
energy neutrino physics, optimizing the performance. We have
restricted ourselves to pressures up to a few bar, at which
electrons above a few 100 keV produce tracks long enough to be identified.\\
First we investigated mixtures dominated by Xenon, having in mind
the search for neutrinoless double beta decay in $^{136}$Xe. We
have looked for additives improving drift properties and gas
amplification. We have concentrated on CF$_{4}$ because it is
cheap, easy to use, non hazardous, and transparent to light. The
latter feature is important if the scintillation light is to be
measured in parallel to the ionization. Other additives can be
considered, for instance
Xe($97$)-CF$_{4}$($2$)-isoC$_{4}$H$_{10}$(1) \cite{vavra,
yprivate}, but to us CF$_{4}$ appears the best compromise. Second,
for experiments on elastic neutrino scattering off electrons, we
have investigated gas mixtures with a large CF$_{4}$ component.
Pure CF$_{4}$ is a high density gas with good drift properties. It
has a relatively low Z, with low multiple scattering making tracks
smoother allowing for a better reconstruction of the electron
direction. We have tried to get around the attachment problem,
which limits the amplification at higher pressures in CF$_{4}$.
Some of the results we obtained are also relevant for dark matter
searches. In these experiments high amplification is required to
detect the tiny fraction of
ionization deposited by a recoiling nucleus. \\
Both Xenon and CF$_{4}$ are good scintillators. This opens the
door to combined measurement of the ionization and the light in a
TPC, which should further enhance the detector performance. In
particular detection of the primary scintillation light may
provide a time reference for the absolute drift time, from which
the absolute event position in the drift direction can be
determined. In the present work however we have concentrated on
the gas amplification. Our results are presented in the following.

\section{Xenon as a double beta decay candidate}
The $^{136}$Xe isotope is a good candidate for the search of
"$0\nu\beta\beta$". The released energy is relatively high
(Q$_{0}=2.479$ MeV), the relative abundance is 8.9$\%$, and it is
relatively easy to enrich. It has a rather high density (5.858
g/l). It has been successfully used in the Gotthard experiment
with a gas TPC \cite{Roland}. Xenon gas acts as the source and the
detector medium. The EXO collaboration is presently investigating
$^{136}$Xe with a liquid TPC, EXO-200 \cite{EXO-200}, having a
much improved sensitivity.  But a gas TPC remains an alternative
for the final version of the EXO experiment, with still superior
sensitivity \cite{EXO}. Xenon has a high ionization yield (W$=22$
eV), it is an efficient scintillator ($175$ nm), it
is mono-atomic and it is relatively easy to purify.\\
In the Gotthard experiment 5\% of CH$_4$ was added as quencher.
The energy resolution and calibration was determined from the
double escape peak of the 2614 keV $^{208}$Tl line, obtained in an
exposure to a $^{232}$Th source illuminating uniformly the gas
volume. The resolution at 1592 keV was found to be
$\sigma(E)/E=3.4 \%$ scaling to $\sigma(E)/E=2.7 \%$ at 2.48 MeV.
This is however a resolution averaged over the entire drift length
(70 cm), since the TPC had no absolute time reference.

\section{CF$_{4}$ for solar neutrino and dark matter search}
CF$_{4}$ has a much lower average Z, and multiple scattering is
smaller than in Xenon. It has a high density (3.72 g/l at 1 bar,
or $1.06\times10^{21}$ e$^{-}$/cm$^{3}$), leading to a compact
detector size. It is more suited for solar or reactor neutrinos.
The high spin of fluorine makes it a good candidate for the
detection of dark matter with spin interaction.  In
addition, CF$_{4}$ is a scintillator, which emits light in the
region from ultraviolet to the visible light \cite{Pansky}. The
primary scintillation photon yield of CF$_{4}$
is about 16($\pm$5)$\%$ of that of Xe as cited in the last reference. \\
Despite its high  electron attachment in high electric fields,
which can cause electron loss, and ageing effects \cite{vavra},
relatively high gas amplification can be achieved \cite{Anderson}.
It was used in several detectors: a gain of $10^{6}$ was reached
with triple-GEM (Gaseous Electron Multiplier) \cite{Sauli1} based
detector in pure CF$_{4}$ \cite{Breskin1}. It was used as filling
gas in the MUNU experiment with a Multi Wire Proportional Chamber
MWPC \cite{Charpak}. It gave good tracking capability, which was
essential in getting results on the electron neutrino magnetic
moment \cite{MUNU1}.

\section{Drift properties and diffusion}
As a first step, we performed some simulations with the Magboltz
\cite{Biagi1} program to study the effect of CF$_{4}$ as additive
in Xe to improve the electron transport properties (drift
velocity, diffusions, etc). The ionization potential of CF$_{4}$
($15.9$ eV) is significantly higher than the excitation energy of
Xe ($8.4$ eV). Comparisons with other quenchers were also made.
Calculations showed that an Xe-CF$_{4}$ mixture has a higher drift
velocity and lower longitudinal and transversal diffusions,
compared with Xe-CO$_{2}$, Xe-isoC$_{4}$H$_{10}$ and Xe-CH$_{4}$.
As mentioned in reference \cite{Christophorou1}, the drift
velocity is high when the electrons are slowed down into an energy
region where the mean scattering cross section $<\sigma_{sc}>$ for
the gas is small, which is the case for the CF$_{4}$ gas.\\ To
give an idea, in Xenon with only 2$\%$ of CF$_{4}$, assuming a
typical drift field of 200 V$\cdot$cm$^{-1}$, the drift velocity
is 0.5 cm$\cdot$$\mu$s$^{-1}$, the lateral diffusion for one cm of
drift is about 225 $\mu$m and the longitudinal diffusion is 285
$\mu$m per cm.\\
We also looked if the properties of CF$_4$ can be improved by an
addition of Xenon.  For CF$_4$ with 2\% of Xenon we find 4.2
cm$\cdot$$\mu$s$^{-1}$ for the drift velocity, 180 $\mu$m per cm
for the lateral diffusion, 178 $\mu$m per cm for the longitudinal
diffusion. A more detailed description of these considerations can
be found in reference \cite{mythesis}. These simulations guided us
throughout our measurements.

\section{Experimental set-up}
All measurements presented in this paper were carried out with a
mini Time Projection Chamber (mTPC) combined with a compact
Micromegas micropattern structure \cite{Giomataris1}. A schematic
view of the mTPC with a Micromegas detection plane is shown in the
left of figure \ref{anodeplane}. A full description of the mTPC
prototype, instrumented with a MWPC, is given in reference
\cite{Carlo}. The drift volume of 10 cm diameter and 20 cm length
was delimited by the cathode at the top and the Micromegas at the
bottom. To achieve a uniform electric drift field, 9 field shaping
rings with 2 cm separation are placed between the cathode and the
Micromegas. They are connected with equal value resistors ($10$
M$\Omega$) and held by spacers made from nonconducting material
(delrin). The Micromegas-mTPC is enclosed in a grounded stainless
steel vessel, which can be evacuated. This prototype was equipped
with various configurations of Micromegas to compare the
performance. In particular, woven wire mesh as well as chemically
etched structures \cite{Patrick} were studied.

\subsection{The Compact Micromegas}\label{micromegas}
\begin{figure}[!ht]
\begin{center}
\begin{tabular}{cc}
\includegraphics[width=8cm,height=6cm]{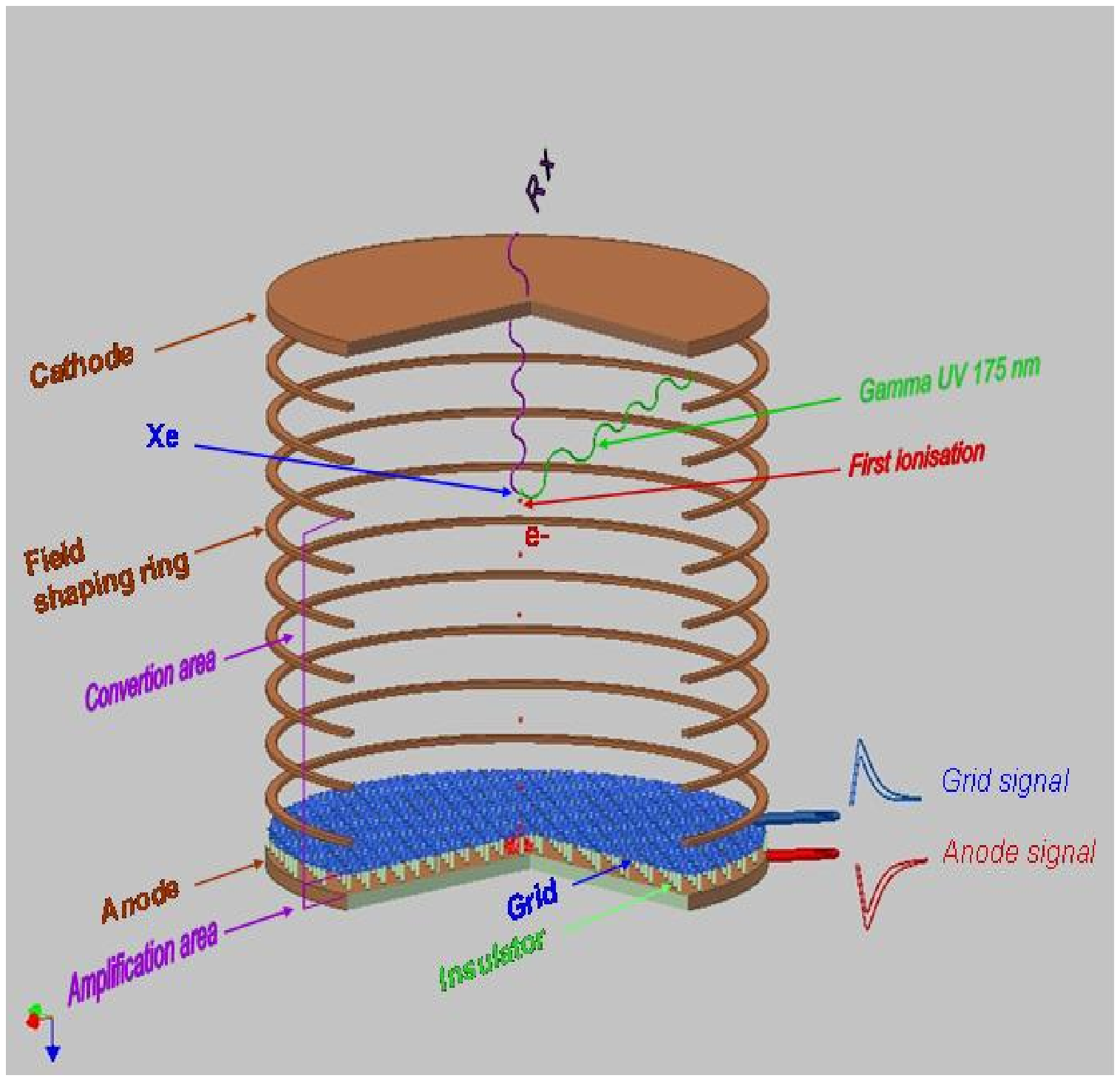}
\includegraphics[width=7.0cm]{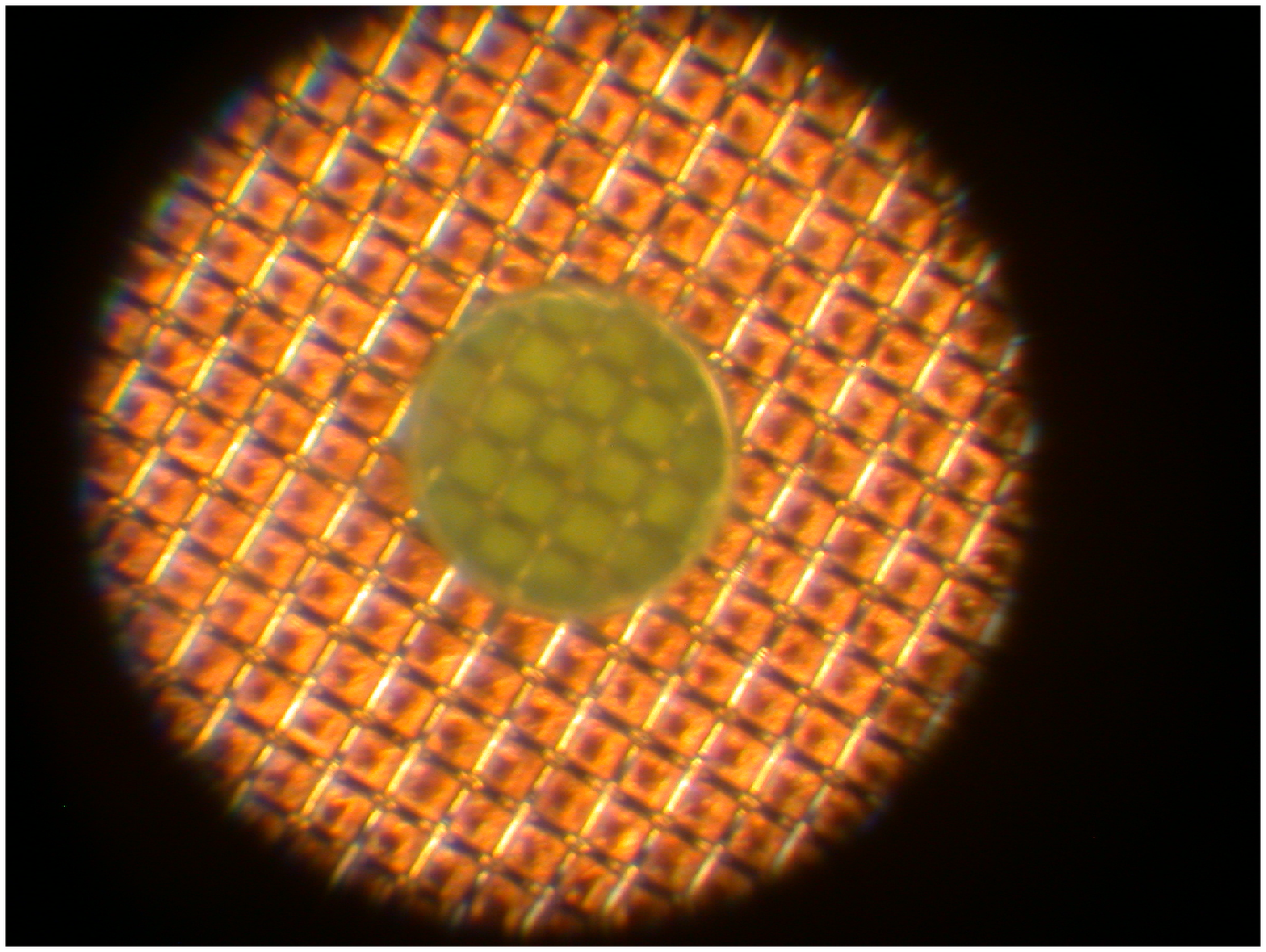}
\end{tabular}
\caption{\label{anodeplane} A schematic view of the
Micromegas-mTPC (left) and a microscopic zoom on a Micromegas grid
with a spacer (right). The diameter of the spacer and the opening
of the grid are $250$ $\mu$m and $53$ $\mu$m, respectively.}
\end{center}
\end{figure}

The compact Micromegas consists of an anode and a grid, compacted
"two in one" and separated with cylindrical pin spacers. The
compact Micromegas is shown in figure \ref{anodeplane}. The anode,
developed at CERN, is a plain copper plated board with 9 cm of
diameter. Kapton spacers with 250 $\mu$m diameter by 100 $\mu$m
height or more and placed every 1 mm, are laid on it by
conventional lithography. The height
defines the amplification gap size.\\
The grid, from the BOPP\footnote{www.bopp.ch} company, is made of
stainless steel wires, with $20$ $\mu$m of diameter, woven with a
pitch of $53$ $\mu$m. It was tensioned delicately and glued with a
thin Araldite layer to the outer part of the anode. The glue layer
is thin enough to have no impact on the gap depth. The grid is
pressed against the spacers.

\begin{figure}[!ht]
\begin{center}
\begin{tabular}{cc}
\includegraphics[width=8cm,height=8cm]{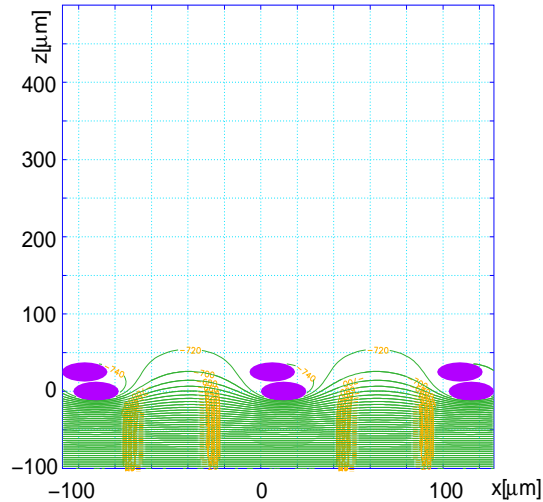}
\end{tabular}
\caption{\label{contv}Electric field configuration near the grid
wires.}
\end{center}
\end{figure}

\noindent This grid is cheap, easy to handle, robust (no damages
were observed on the grid even after high voltage discharges). \\
Best results with good charge collection have been obtained for a
gap around $100$ $\mu$m \cite{Giomataris2}, which justifies our
choice of this height for all measurements at 1 bar.
Field non uniformities, including the drift volume (20 cm) and the
amplification gap (100 $\mu$m), are studied with the Garfield
\cite{Garfield} program as shown in figure \ref{contv}. The $2$D
configuration of Garfield was used.

\noindent In reality, the grid has a $3$D structure, but it can be
locally approximated to good precision by a $2$D structure. The
$3$D crossed wires in the micromesh grid are approximated by
superposed circles, with $20$ $\mu$m diameter. One of them is
slightly displaced. The wires are repeated every $53$ $\mu$m.\\
The electric field is homogeneous in both the conversion and the
amplification region, where the field is highest, which is
required for good gain uniformity.  There is a bulge in the
intermediate opening of the microgrid, which is however of no concern.\\
Great care has been taken when handling the detection plane, to
avoid any gap deformation. Obviously, defects of flatness are
the source of gain fluctuations and therefore affect the energy
resolution. In all of our tests, the ratio between the
amplification and the drift electric fields exceeds $250$ for all
gas mixtures. This value is sufficient to permit a full electron
transmission, with intrinsic ion feedback
suppression, leading to nearly total charge collection.\\
Before installation, the compact Micromegas is washed with
isopropyl alcohol and dried with a thin vacuum cleaner brush to
remove dusts and scraps from manufacturing and to keep humidity
low. Traces of humidity in this narrow gap can adversely affect
the energy resolution, while dust particles lead to discharges.

\subsection{Electronics and gas systems}
The detector was operated with the cathode at a negative
potential, the anode at ground and the grid at a negative
potential of several hundred Volts.\\
The signal is taken from the grid. It is collected on a capacitor
(C$=10$ pF), amplified by a charge-sensitive preamplifier and
amplified again and shaped in a gaussian form by a spectroscopic
amplifier.
A multichannel analyzer sorts the incoming pulses by their height.
The mean value of the current is measured with a slow base time
oscilloscope via a nanoampermeter.\\
To achieve good purity, the chamber was evacuated to $10^{-6}$
mbar while heating before each gas filling. The mixing of Xe and
CF$_{4}$ at ambient temperature was done by controlling the
partial pressure within the chamber itself.\\
After filling, the gas was circulated continuously through a
purification system. It includes two "\emph{oxysorb}" filters: one
to remove electronegative impurities such as O$_{2}$ and H$_{2}$O
and the second one in series to monitor the purity. A cold trap
maintained at a temperature fluctuating around $-109 ^{\circ}$C
serves to remove water and possible remaining freon
contaminations. Measurements started after one day of gas
circulation. Moreover, at regular intervals, the gas composition
is controlled with a mass spectrometer mounted near the chamber,
analyzing the inflowing and the outflowing gas.

\section{Main results}
Charge spectra from the grid signal are recorded for each gas
mixture with different proportions of the additive, exposing the
gas volume to a $^{55}$Fe source (5.9 KeV X-rays). The source was
placed on the cathode. All measurements are done under the same
conditions.
The acquisition time is $100$ s for all measurements.\\
In this section we report on results obtained at atmospheric
pressure for various Xe-CF$_{4}$ admixtures. The drift field has
been maintained constant during all measurements at $200$
V$\cdot$cm$^{-1}\cdot$bar$^{-1}$. \\
The main emphasis was put on the energy resolution of the detector
at low energy and the determination of the gas gain. These factors
were determined in the following way:
\begin{itemize}
    \item The effective gain of each gas mixture was measured by comparing
the peak of the pulse height distribution generated by the
$^{55}$Fe X-ray source with that of a calibrated charge deposited
at the test input of the charge sensitive
preamplifier.\\
The charge collected on the grid represents the total number of
ions created after the amplification and is equal to $CV$, where
$C$ is the input capacitance of the preamplifier and $V$ is the
voltage at the end of the collection, determined by comparison
with the calibration pulses. Thus, the gas gain is given by the
relation:
\begin{equation}
G=\frac{CV}{eE/W_{i}},
\end{equation}
where $e$ is the elementary charge, $E$ is the deposited energy of
the incident photon (equal to 5.898 KeV in the case of the
$^{55}$Fe source) and $W_{i}$ is the average energy required to
ionize the gas (54 and 22 eV for CF$_{4}$ and Xe respectively).
\item The relative intrinsic FWHM energy resolution at half height
is given by:
\begin{equation}\label{Eexp}
R_{E}=2.35\frac{\sigma(E)}{E},
\end{equation}
where $\sigma(E)$ is the standard deviation and $E$ is the average deposited energy.\\
In these measurements, the electronic noise only leads to a small
increase of the total resolution and was neglected.
\end{itemize}

\subsection{Charge collection in Xenon gas}\label{favorxe}
\begin{figure}[!ht]
  \begin{center}
   \includegraphics[width=8cm, height=6cm]{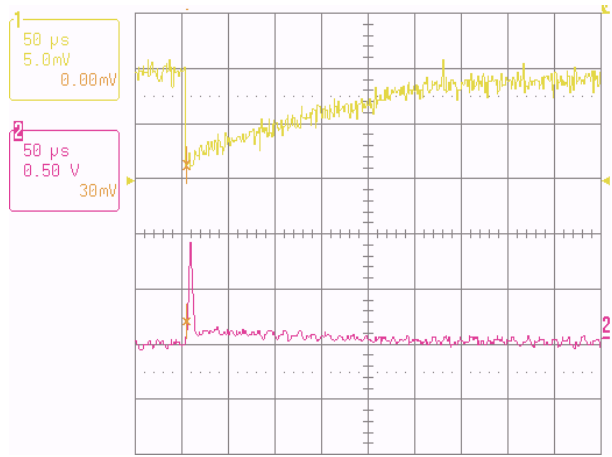}\\
  \caption{The shape of the preamplifier output (top) and the
  spectroscopic amplifier (bottom), recorded in the oscilloscope for Xe($98$)-CF$_{4}$($2$)
   gas mixture at $1.00$ atm and $-470$ V on the grid voltage.}\label{eventxe98cf4}
  \end{center}
\end{figure}

We have done experimental studies of the gas gain of Xe-CF$_{4}$
mixtures, going from 2$\%$ of CF$_{4}$ in 98$\%$ of Xe to
(50-50)$\%$ by volume. A comparison with 2$\%$ of
IsoC$_{4}$H$_{10}$ in Xenon, often given as the best quencher for
Xe, was also performed.
No amplification was obtained with pure Xe.\\
A typical preamplifier response to $^{55}$Fe $5.9$ keV X-rays in
Xe(98)-CF$_{4}$(2) gas at $1.00$ atm, is shown in figure
\ref{eventxe98cf4}. The shaping of the signal allows to catch the
entire induced charge. The integrated signal current measured with
a nano-ampermeter  is very small (few hundreds pA).

\begin{figure}[!ht]
\begin{center}
\begin{tabular}{cc}
\includegraphics[width=7.6cm,height=9cm]{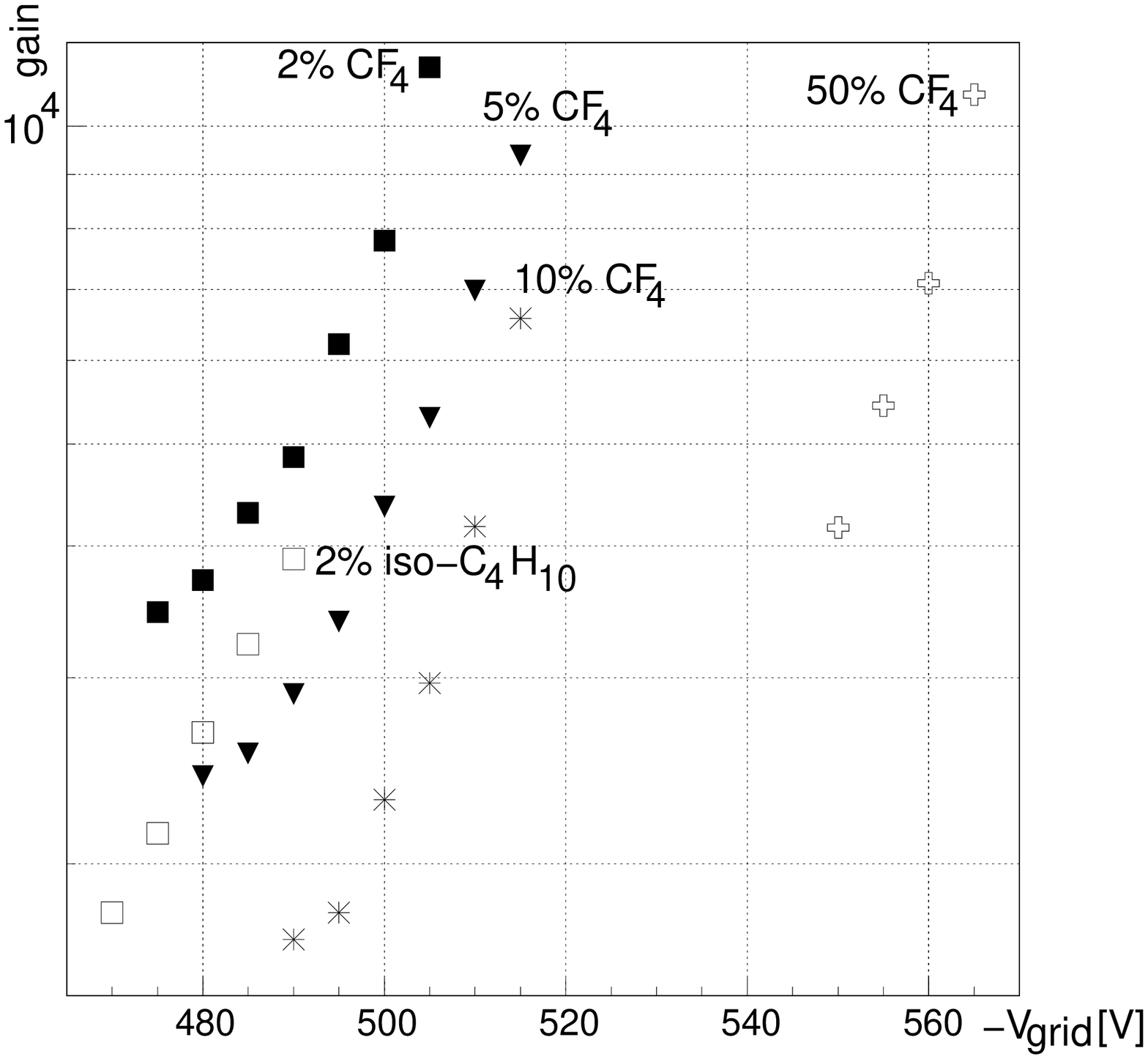}
\includegraphics[width=7.6cm,height=9cm]{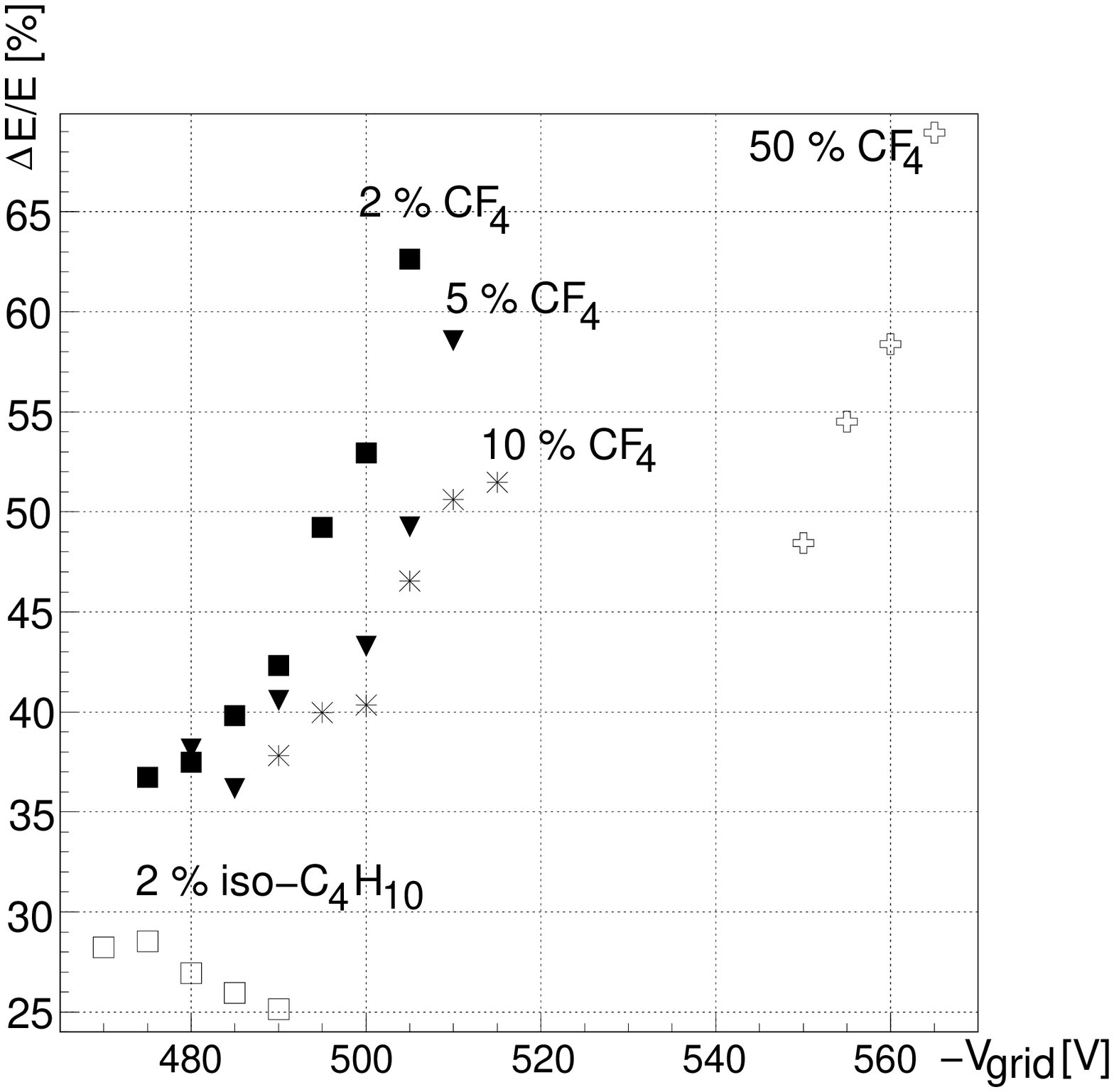}
\end{tabular}
\caption{Gas gain G measurements (left) and $^{55}$Fe energy
resolutions R$_{E}$ (right) versus grid voltages, with Xe-CF$_{4}$
(full markers) and Xe-IsoC$_{4}$H$_{10}$ ($\square$) admixtures at
atmospheric pressure and $100$ $\mu$m gap.}\label{Xe-CFiso}
\end{center}
\end{figure}

\noindent We summarized in figure \ref{Xe-CFiso} the effective gas
gain curves and energy resolutions at $6$ keV of energy, obtained
in various gas mixtures as a function of the grid voltage. The
gain was measured for increasing grid-anode voltages, up to
the point where discharges occur. \\
For a given gain, the grid voltage must be increased at higher
CF$_{4}$ admixtures. Usually, the addition of the quencher gas is
meant to decrease the operating voltage and to increase the gain
amplification, which is not the case here. \\
The observed gas gains are large enough to allow the detection of
small signals in the ionization mode of the Micromegas-mTPC. They
exceeded $10^{3}$ for all Xe-CF$_{4}$ gas mixtures investigated. A
remarkable fact is that we obtain the best gain for a given
voltage with only 2$\%$ of CF$_{4}$.

\begin{figure}[!ht]
\begin{center}
\begin{tabular}{cc}
\includegraphics[width=7cm,height=8cm]{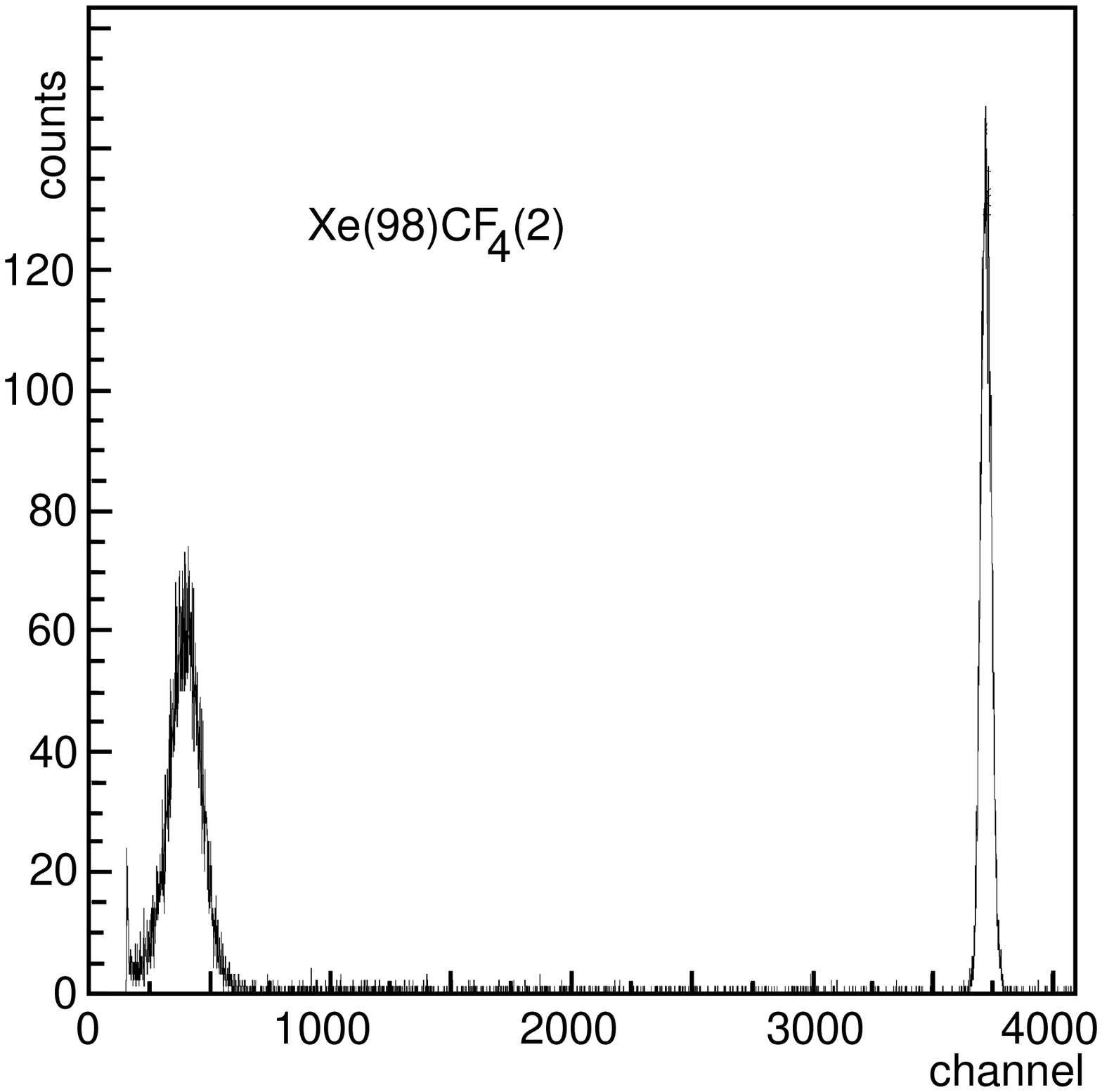}
%\hspace{0.2cm} & \hspace{0.2cm}
\includegraphics[width=7cm,height=8cm]{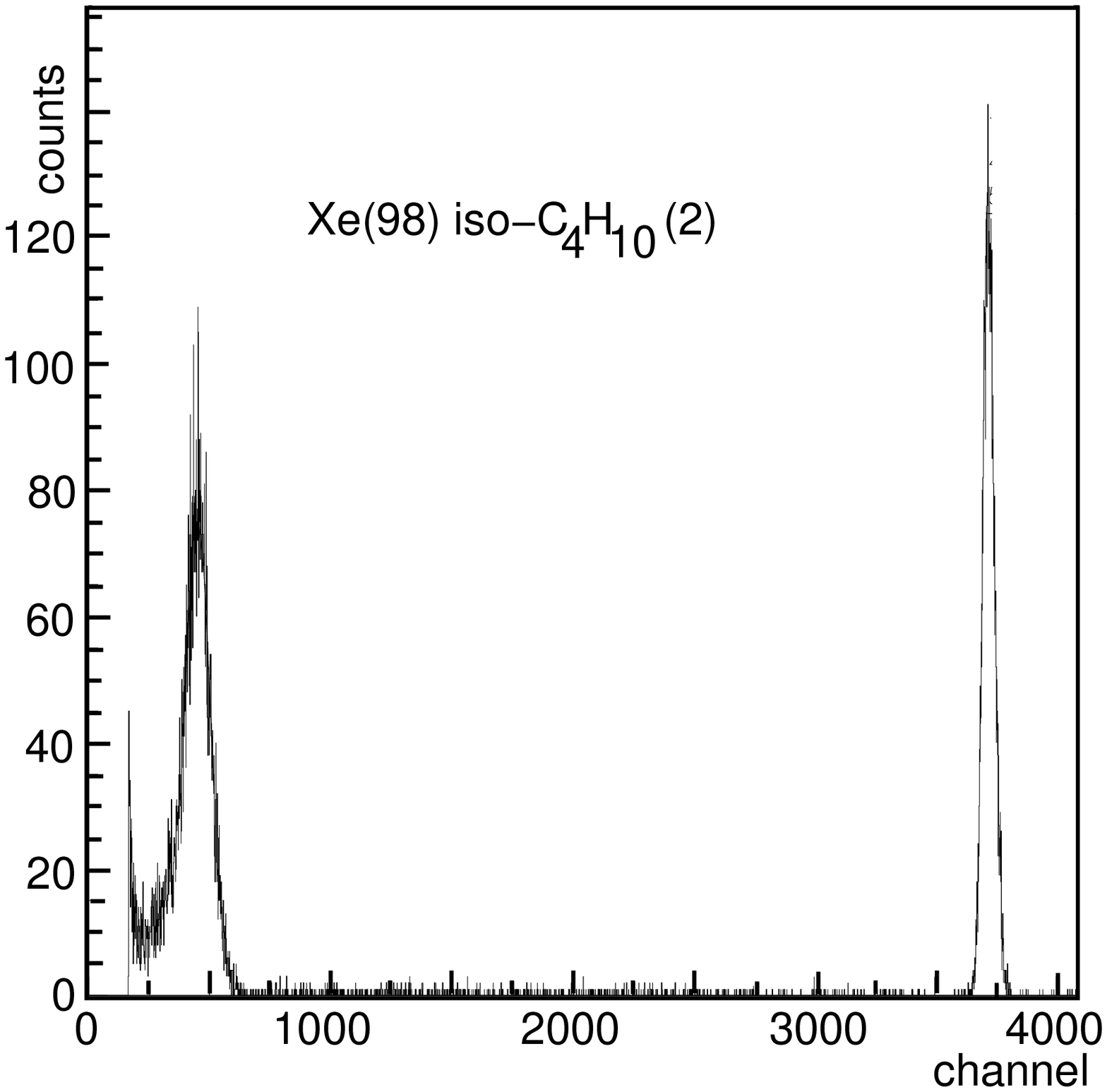}
\end{tabular}
\caption{\label{compcfiso} The response to the 5.9 keV X-rays from
a $^{55}$Fe source with a compact Micromegas in 1 bar of
Xe-CF$_{4}$ (left) and in Xe-isoC$_{4}$H$_{10}$ (right) gas
mixtures at the same proportion (2$\%$) and the same settings. The
contribution of the electronics noise to the energy resolution is
given by the pulser peak width to the right.}
\end{center}
\end{figure}

\noindent The half-width energy resolution of
Xe($98$)-CF$_{4}$($2$) admixtures is in the range of ($35$ -
$65$)$\%$ in the working grid voltage region, compared with ($25$
- $29$)$\%$ with Xe(98)-isoC$_{4}$H$_{10}$(2) (see the right of
figure \ref{Xe-CFiso}). However, comparing with the same
proportion of isoC$_{4}$H$_{10}$ quencher added to Xe, the
Xe($98$)-CF$_{4}$($2$) admixture gives higher gas gain ($1.4$
times better) at the same grid voltage (V$_{g}=-480$ V) with same
settings and conditions, as shown in figure \ref{compcfiso}.\\
Even under the worst case of the Fano factor F$=1$, the energy
resolution is not as good as expected
$\big{(}$$R_{E}=2.35\sqrt{\frac{FW_{i}}{E}}$\big{)} with CF$_{4}$
addition. This worsening of the energy resolution can be explained
by the strong contribution of the
electron attachment in CF$_{4}$ \cite{Christophorou1}.\\
To conclude this section, we can reach high gains at lower applied
voltages, with only 2$\%$ of CF$_{4}$ added to Xe (see figure
\ref{Xe-CFiso}). The Xe($98$)-CF$_{4}$($2$) gas mixture is well
suited for a search of neutrinoless double beta decay in
$^{136}$Xe as a double beta candidate for its good charge
collection and good transport properties. The lower gain at higher
percentage of CF$_{4}$, can be explained by the loss of electrons,
which inhibits the ionization in the avalanche region. Ageing of
the gas was observed and found to be less at low
CF$_{4}$ concentrations \cite{vavra}.\\
Large and flat (with a small drift volume) Micromegas-TPC filled
with Xe(98)-CF$_{4}$(2) at 4-5 bar of pressure could be convenient
for medical imaging and safety control in ports and airports. It
may be a good alternative to existing schemes with MSGC's
\cite{Gobbi} and with MWPC's \cite{WC-conf}. Micromegas-based
gaseous photomultipliers filled with Xe-CF$_{4}$ admixture could
be developed, for the UV and the visible spectral range. Progress
in this field with Multi-GEM micropattern cited in reference
\cite{Chechik} are very encouraging.

\subsection{Reducing the CF$_{4}$ attachment}\label{cf4attachment}
Pure CF$_{4}$ was used in the MUNU TPC, and its drifting and gain
properties are well known. Gas multiplication was made around thin
wires. As mentioned one problem is the electron attachment
occurring at higher fields in the avalanche region, which leads to
electron losses, gain reduction, and to a deterioration of the
energy resolution. In the subsequent studies, we estimated the
electron loss at different CF$_{4}$ gas pressures with Magboltz
and Imonte \cite{Biagi1} interfaced to Garfield, under the exact
experimental conditions (pressure, temperature, voltages ...).\\
The electron loss increases with the gas pressure, which is
confirmed by experimental results. At $1$ bar of pressure, the
high electric field (about $400$ kV/cm) around the anode wires,
causes a serious electron loss and the charge collection
efficiency is about 13$\%$. At $3$ bar of CF$_{4}$, only 5$\%$ of
the electrons survive. This decrease of the collection efficiency
contributes to the worsening of the energy resolution. More
details for electrons loss
calculations are in reference \cite{mythesis}.\\
A comparison between energy resolutions of the MUNU-TPC for $1$
and $3$ bar is mentioned in reference \cite{Zori}, confirming that
the energy resolution at $1$ bar is about $1.7-2$ times
better than at $3$ bar.\\
We decided to test whether replacing the MWPC used in MUNU by a
Micromegas structure helps in reducing the attachment. \\
A comparison between $^{55}$Fe pulse height spectra of the compact
Micromegas \cite{mythesis} and the MWPC \cite{Carlo} in pure
CF$_{4}$ at 1 bar of pressure and same conditions and settings,
gave the Micromegas a clear advantage. The energy resolution is
improved by a factor 1.3. A first advantage of the Micromegas
structure is that the electric field near the anode is very
homogenous. In the case of the MWPC, the electric field around the
anode wire is very high compared with the amplification field in
the Micromegas gap region. The attachment is therefore less with
the Micromegas.\\
We further considered an additive to improve the properties of
CF$_{4}$. Xe and Ar gases are good candidates to quench CF$_{4}$
gas. The first one is simply chosen because of its low operating
voltage, and the second one has zero attachment.\\
The first ionization potential of Xe ($12.13$ eV), compared with
the excitation energy of CF$_{4}$ ($12.5$ eV) molecules, allows
the addition of Xe to CF$_{4}$. The high electric field in the
amplification region leads to the following decomposition of
molecules \cite{Turban}:
\begin{eqnarray}
e^{-} + \hbox{CF}_{4} \rightarrow (\hbox{CF}^{*}_{3})^{+} +
\hbox{F}^{-} + 2e^{-}\\
e^{-} + \hbox{CF}^{+}_{4} \rightarrow (\hbox{CF}^{*}_{4})^{+} +
e^{-}
\end{eqnarray}
When a stable state is reached, dissociated fragments
((CF$^{*}_{3}$)$^{+}$ and (CF$^{*}_{4}$)$^{+}$) emit photons in
the UV (peaked at 160 nm) and the visible range ($\lambda=620$ nm)
\cite{Zhang}. Light signals were also recorded
by the MUNU photomultipliers \cite{Avenier}.\\
Measurements are carried out with the same prototype described
above using parameters similar to those for the charge collection
in Xe study (pressure, amplification height, drift field...).

\begin{figure}[!ht]
\begin{center}
\begin{tabular}{cc}
\includegraphics[width=7.6cm,height=8cm]{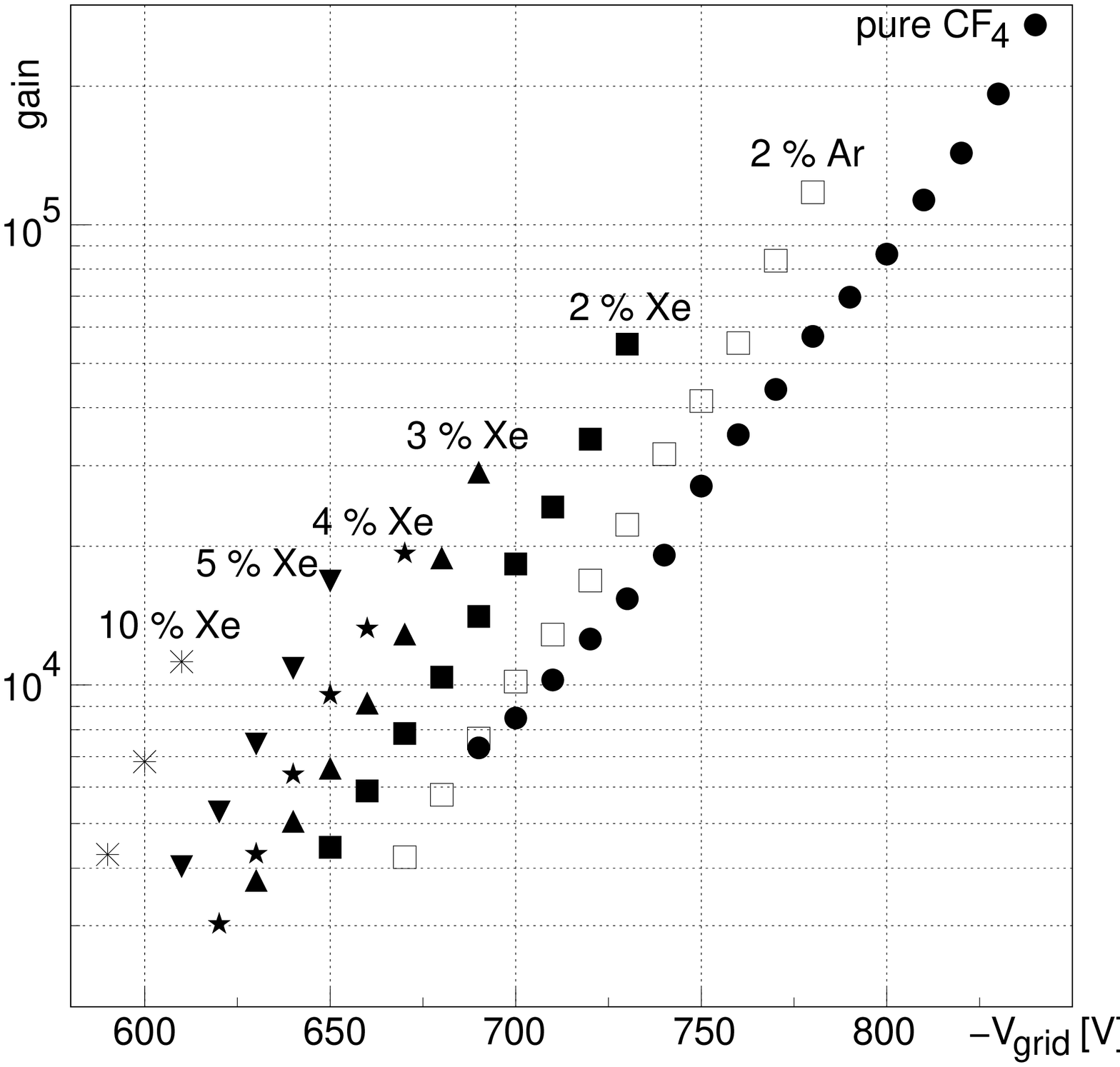}
\includegraphics[width=7.6cm,height=8cm]{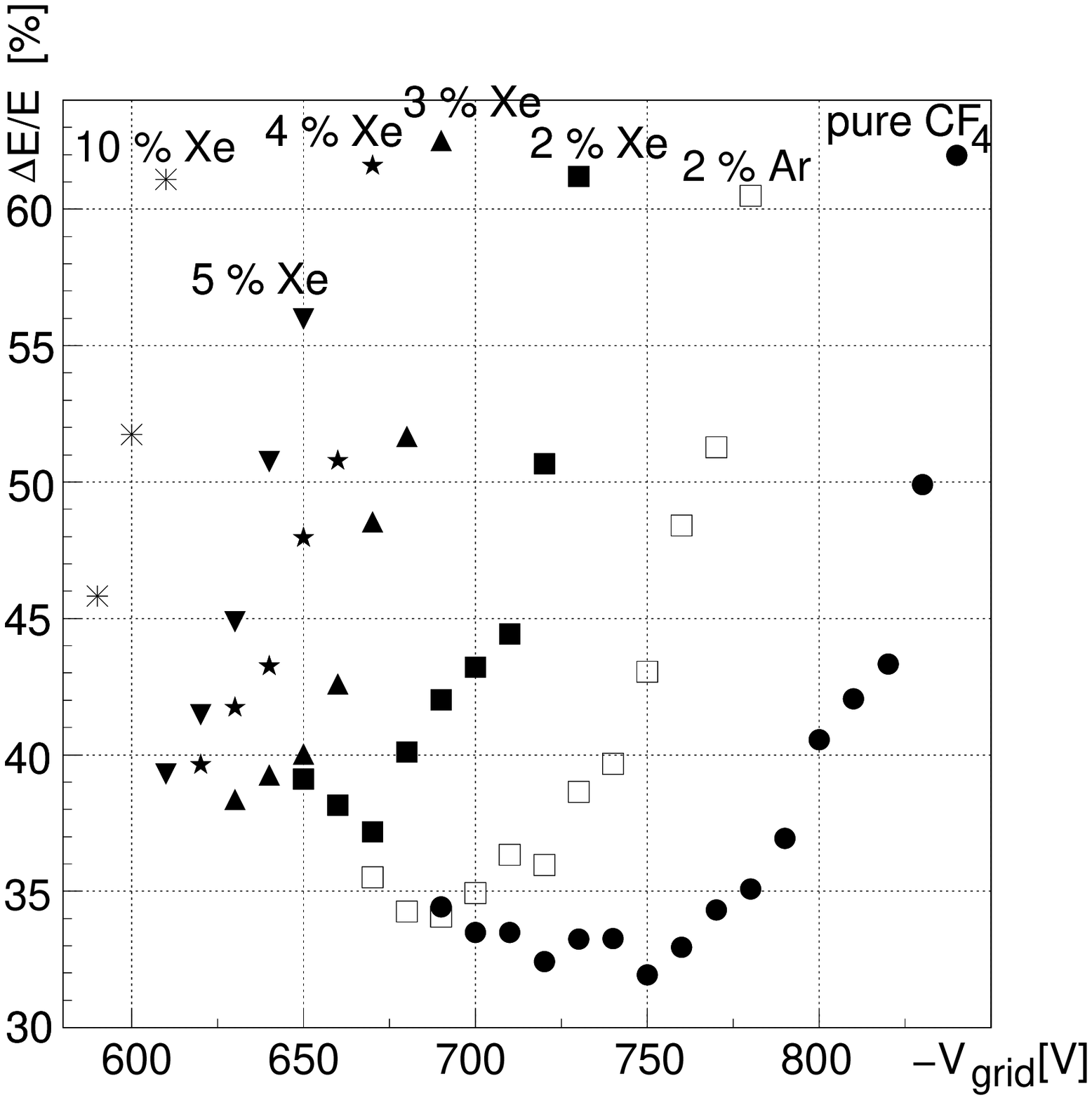}
\end{tabular}
\caption{Gas gains G (left) and $^{55}$Fe energy resolutions
R$_{E}$ (right) measurements with CF$_{4}$-Xe (full markers) and
CF$_{4}$-Ar ($\square$) admixtures from pure CF$_{4}$ to 10$\%$ of
Xe and 2$\%$ of Ar quenchers at atmospheric
pressure.}\label{CF-XeAr}
\end{center}
\end{figure}

\noindent Results of our measurements are summarized in figure
\ref{CF-XeAr}. In pure CF$_{4}$ gas, the working region is around
$150$ V grid voltage. Adding noble gas decreases the maximum
possible gain.
It increases the gas gain at a given low operating voltages.\\
Moreover, this addition decreases the probability of photoeffect
on the grid wall, caused by ultraviolet photons from dissociated
fragments of CF$_{4}$. The presence of Xe atoms can prevent or
more precisely delay the formation of those fragments, especially
dissociative processes, because the operating voltage is lower
than before. \\
A small addition of Xe reduces the working voltage, and increases
the gas amplification by more than 10 times. At the same grid
voltage, we obtain a gas gain $10$ times greater with
CF$_{4}$($98$)-Xe($2$) than with pure CF$_{4}$. This can be
explained by the decrease of the number of electrons lost in the
avalanche ionization region. At higher pressure the high
attachment of CF$_{4}$ makes the Micromegas less efficient. \\
As presented in the right plot of figure \ref{CF-XeAr}, the energy
resolution is not improved when adding Xe or Ar quenchers. It is a
consequence of the increase of diffusions, the lowering of the
drift velocity and the photoeffect of Xe and Ar in these
admixtures. But the lowering of the operating voltages and the
higher gains obtained confirmed the good collection efficiency.

\section{Use in large size devices for rare event detection}
In the envisioned rare event experiments large detector masses are
required and background considerations are important. In this
context compact Micromegas look promising. They are simple, and
can be made from materials with good radiopurity. Mesh grids can
be obtained in large sizes ($2.5 \times 30.5$ m$^{2}$). A large
size Micromegas (50 cm of diameter, otherwise similar to the
structures described so far)\footnote{Compass detections planes:
GEM and Micromegas are about ($31\times31$) and ($40\times40$)
cm$^{2}$.} was tested in the Gotthard-TPC \cite{Roland} with 1 bar
of P10 (Ar(90)CH$_{4}$(10)) and CF$_{4}$ gases. The drift distance
was 67 cm and the drift field 200 V/cm. Detailed results are given
in reference \cite{parisconf}.

\begin{figure}[ht]
\begin{center}
\includegraphics[width=7cm,height=7cm]{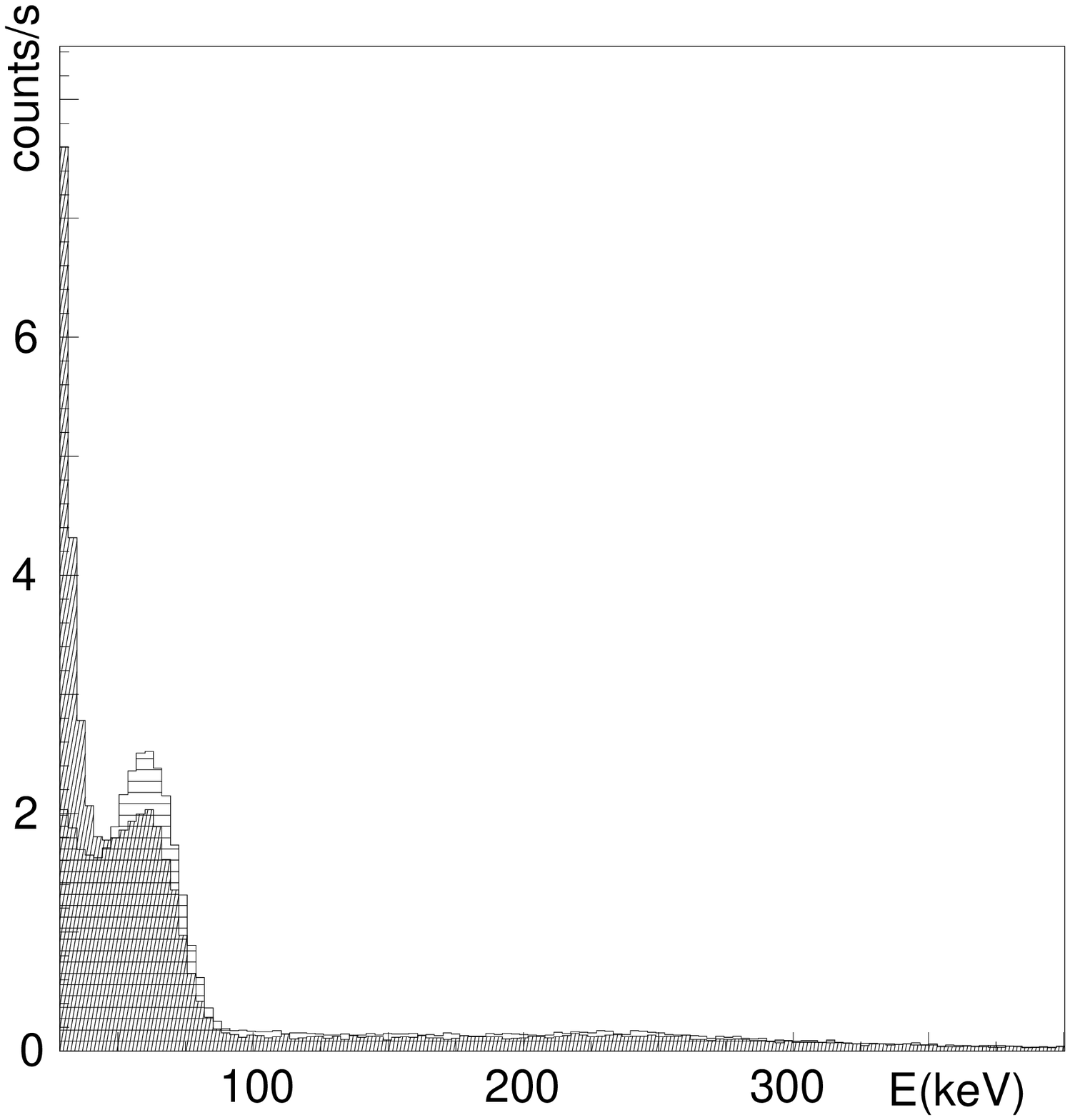}
\includegraphics[width=7cm,height=7cm]{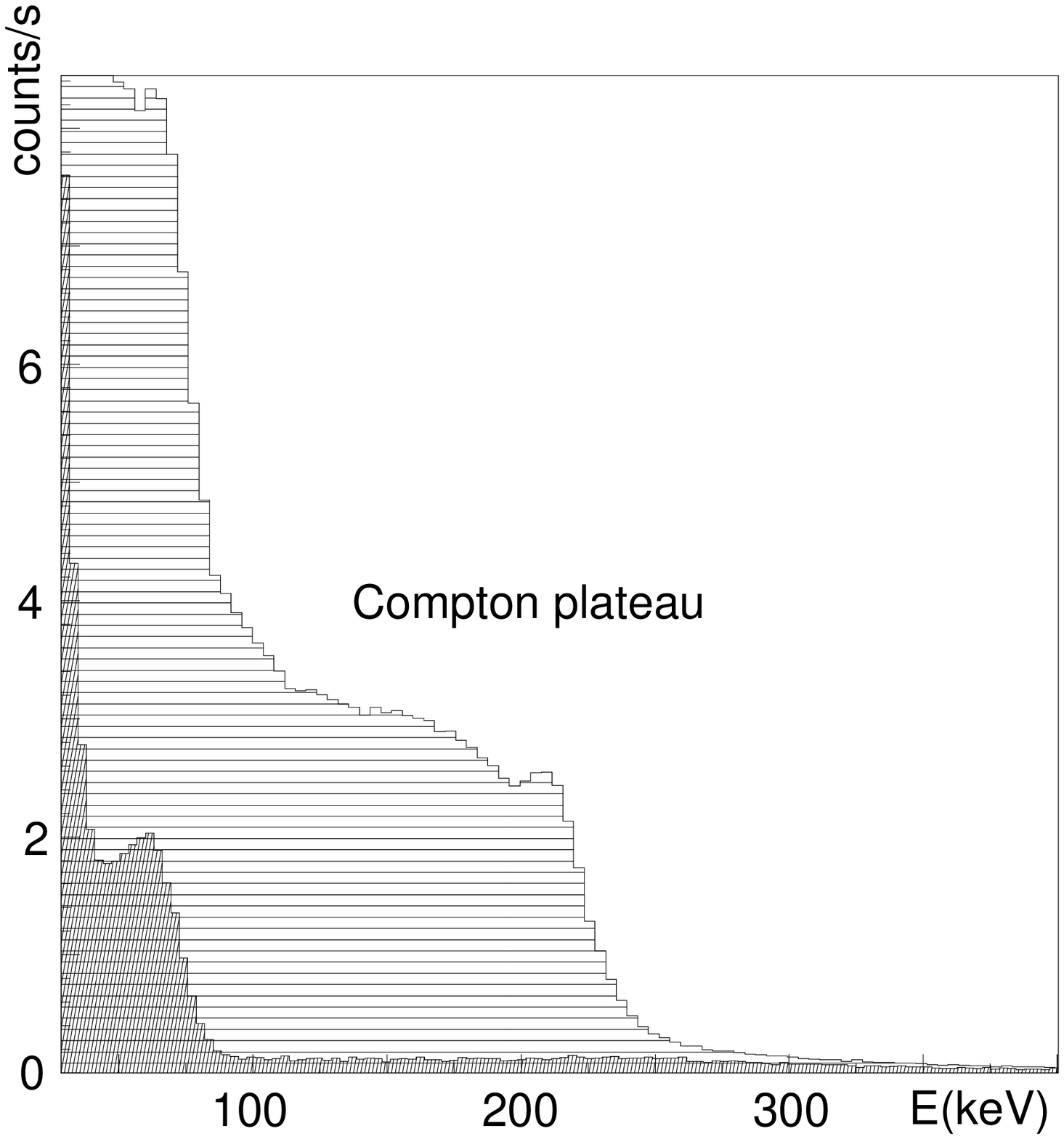}
\caption{\label{50cm-amintambaext}The response to the internal
$^{241}$Am source and the external $^{241}$Am (left) and
$^{133}$Ba (right) sources, with the $50$ cm diameter Micromegas
in the Gotthard TPC with $1.03$ atm of CF$_{4}$. The cathode is
set at -13.8 kV.}
\end{center}
\end{figure}

\noindent The energy calibration was performed with an internal
$^{241}$Am source (37 kBq) and two external gamma ray sources
($^{241}$Am and $^{133}$Ba with $370$ kBq activity each). The
spectra with the internal source alone and, in addition, the
external sources, are shown in figure \ref{50cm-amintambaext}.\\
The Compton edge at 207 keV and 228 keV of the Ba lines at 356 and
382 keV is clearly visible in the right plot of the figure, as
well as the $81$ keV full energy peak. Better containment with the
larger Micromegas is thus obvious. The energy resolution is 10$\%$
 FWHM at 228 keV, averaged over the Micromegas.\\
Moreover to have a large target or source mass without
unmanageable volume requires working at high pressure. Reasonable
electron tracks, with identification of the increased ionization
at the end, can be observed up to say 10 bar of pressure in a TPC.
All of our measurements reported so far were performed at a
pressure of 1 bar. But we varied the pressure and found that the
maximal achievable gain decreases with pressure.  The same effect
was observed with a triple-GEM detector \cite{Orthen, Bondar1},
operated at different pressures of argon, krypton and Xenon. We
have shown however that it is possible to work at higher pressure
if the Micromegas gap is increased.

\begin{figure}[ht]
  \begin{center}
   \includegraphics[width=26pc,height=24pc]{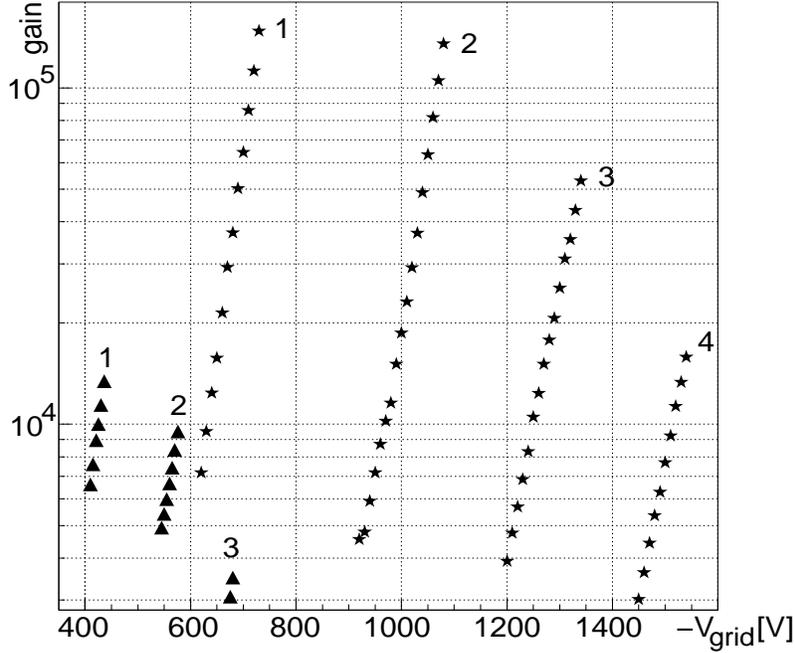}\\
  \caption{Gas gains measured in P$10$ gas, with $75$ $\mu$m ($\blacktriangle$)
  and $225$ $\mu$m ($\star$) gap heights at different pressures as a function
  of the grid voltage, using a $^{55}$Fe source.}\label{p10gapeffect}
  \end{center}
\end{figure}

\noindent We have studied the dependence of the gas gain of the
Micromegas detector as a function of gap. Two different gaps were
investigated: $75$ and $225$ $\mu$m at various gas pressures
extending from $1$ bar to the highest achieved pressure 4 bar.
Figure \ref{p10gapeffect} shows measurements of the gas gain
versus the grid potential obtained for P$10$ at ($1, 2, 3$) bar
for $75$ $\mu$m and ($1, 2, 3, 4$) bar for $225$ $\mu$m gaps. With
the larger gap value ($225$ $\mu$m), gas gains greater than
$10^{4}$ are comfortably achievable.  The collection efficiency is
higher with the $75$ $\mu$m gap, but the dynamic range is smaller.
One can reach the same gain as with a higher gap with a lower
applied voltage. With a $225$ $\mu$m gap, much higher voltages are
needed to reach the same gain, but the working region is much
larger, and much higher gains are achievable. This makes it
possible to work at higher pressure. Moreover with higher gaps,
the relative gap variations over the entire area is reduced. This
insures better gain uniformity in the avalanche over larger
detection plane surfaces.\\
The energy resolution at 6 keV is slightly better at lower
pressure of P$10$ (22$\%$ and 26$\%$ at 1 and 4 bar respectively),
with approximately the same gains. This is not the case for
CF$_{4}$, where the energy resolution is remarkably worse when
increasing the pressure. The energy resolution at $1$ bar (36$\%$)
is about $1.5$ times better than at $2$ bar (54$\%$) \cite{mythesis}.\\
The Micromegas with 225 $\mu$m gap was also successfully operated
in Xe(98)CF$_{4}$(2) at pressures up to 4 bar. Gain of $10^{3}$
and energy resolution of 50$\%$ at 30 keV have been achieved at 4
bar of pressure \cite{parisconf}.
\section{Conclusions}
TPCs have shown to be powerful instruments in low energy, low
count rate, neutrino physics. A tight event selection is possible
exploiting the imaging capability of these devices. We have
investigated the properties of various gas mixtures in view of
future experiments, amplifying the charge in Micromegas
structures, which are reliable and suited for large size devices.
For double decay searches, we conclude that a mixture of Xenon
(enriched in $^{136}$Xe) with a few percent of CF$_4$ is an
attractive possibility. It gives a large drift velocity, and
allows to reach high gas amplification, essential factors for a
successful experiment. For the study of neutrino electron
scattering, where it is important to have a light gas to minimize
multiple scattering, we showed that CF$_4$, with a few percent of
Xenon reducing the attachment, also gives high gains. All these
gas mixtures are transparent to light, so that a simultaneous
measurements of the scintillation is possible. Neutrino
experiments require large detector masses. To avoid excessive
dimensions, it is necessary to work at pressures of a few bar.
This can be done, with the Micromegas, by suitably adapting the
amplification gap. In a next phase we want to investigate more
precisely the spatial resolution which can be achieved in gas TPCs
filled with these gas mixtures, as well as
the energy resolution at energies up to 2 MeV.\\

\acknowledgments The results cited in this paper are part of
L.Ounalli PhD work in Neuch{\^a}tel. This work was partially
supported by the "Fonds National Suisse". The authors would like
to express their deep gratitude to V. Zacek, J. Busto Y.
Giomataris and F. Juget for their precious comments and their
suggestions concerning this research. The authors are also
thankful to members of the EXO collaboration for stimulating
discussions, and to R. Oliveira (CERN) for providing the anodes
with the spacers.

%%%%%%%-------------------------------------------------------

%%%%%%%-------------------------------------------------------

\end{document}